\documentclass{llncs}
\usepackage{ifthen}

\newcommand{\predref}[2]{{\bf #1/#2}}
            \def\postscriptfig{\@ifnextchar[{\@scaledpostscriptfig}{\@postscript
fig}}
            \def\@scaledpostscriptfig[#1]#2#3{%
        \begin{figure}%
            \centerline{\includegraphics[#1]{#2}}
            \caption{#3}
            \label{fig:#2}
        \end{figure}}
\newcommand{\secref}[1]{Sect.~\ref{sec:#1}}
\newcommand{\tabref}[1]{Tab.~\ref{tab:#1}}

\newcommand{\Tabref}[1]{Table~\ref{tab:#1}}
\newcommand{\program}[1]{{\sf #1}}

\newcommand{\const}[1]{\texttt{#1}}
\newcommand{\file}[1]{\texttt{#1}}

\newcommand{\reffont}{\bf}
\newcommand{\urlref}[1]{\footnote{\url{#1}}}
\def\term{}
\renewcommand{\term}[2]{%
        \ifthenelse{\equal{\protect}{\protect#2}}{%
            {\reffont #1}}{%
            {\reffont #1}({\it #2})}}
\renewcommand{\arg}[1]{\ifmmode\mbox{\em #1}\else{\em #1}\fi}
\newcommand{\bnfmeta}[1]{\ifmmode{\langle\mbox{\it #1}\rangle}\else$\langle\mbox{\it #1}\rangle$\fi}

\begin{document}

\title{Portability of Prolog programs: \\
       theory and case-studies}
\titlerunning{Portability of Prolog programs}

\author{Jan Wielemaker\inst{1} and V\'{\i}tor Santos Costa\inst{2}}
\authorrunning{Wielemaker and Santos Costa}
\institute{VU University Amsterdam, The Netherlands, \\
	   \email{J.Wielemaker@cs.vu.nl}
	   \and
	   DCC-FCUP \& CRACS-INESC Porto LA\\
	   Universidade do Porto, Portugal\\
	   \email{vsc@dcc.fc.up.pt} }

\maketitle

\begin{abstract}
(Non-)portability of Prolog programs is widely considered as an
important factor in the lack of acceptance of the language. Since 1995,
the core of the language is covered by the ISO standard 13211-1. Since
2007, YAP and SWI-Prolog have established a basic compatibility
framework. This article describes and evaluates this framework. The aim
of the framework is running the same code on both systems rather than
migrating an application. We show that today, the portability within the
family of Edinburgh/Quintus derived Prolog implementations is good enough
to allow for maintaining portable real-world applications.
\end{abstract}

%\tableofcontents

%================================================================
\section{Introduction}

Prolog is an old language with a long history, and its user community
has seen a large number of implementations that evolved largely
independently. This situation is totally different from more recent
languages, e.g, Java, Python or Perl. These language either have a
single implementation (Python, Perl) or are controlled centrally (a
language can only be called Java if it satisfies certain
standards~\cite{java:JCP}).  The Prolog world knows dialects that are
radically different, even with different syntax and different
semantics (e.g., Visual Prolog~\cite{visual-prolog}). Arguably, this is
a handicap for the language because every publically available
significant piece of code must be carefully examined for portability
issues before it can be applied.  As an anecdotal example, answers to
questions on \emph{comp.lang.prolog} typically include ``on Prolog
XYZ, this can be done using \ldots'' or ``which Prolog implementation are you
using?''.

In this work we will investigate portability issues in a number of
modern Prolog implementations. We shall use systems that implement the
ISO standard to a large
extend~\cite{bagnara:99a,DBLP:conf/iclp/SzaboS06}. We remark that,
although any program larger than a few pages requires modularity, the
ISO standard for modules was never accepted by most Prolog developers.
To address this problem, we restrict ourselves to Prolog systems that
implement a module system descending from the Quintus module system.
This includes Quintus Prolog itself~\cite{QUINTUS:manual}, SICStus
Prolog~\cite{SICStus:manual}, Ciao~\cite{ciao},
SWI-Prolog~\cite{SWI-Prolog:manual}, and YAP~\cite{yap}. We further
assume that all target systems provide a term-expansion facility (a
macro-facility inherited from the Edinburgh tradition), a second
requirement for our approach.

%================================================================
\section{Portability approaches and related work}
\label{sec:approach}

Software portability is a problem since the day the second computer
was built. In the case of Prolog, we expect that at least basic
portability requirements are fulfilled: there are few syntactic
incompatibilities, and the core language primitives have to a large
extent the same semantics. This is the case for the family of
implementation that is subject in this study. Beyond that, the
implementations vary widely; notably in \textbf{(i)} the organisation
of the libraries; \textbf{(ii)} available library primitives; and
\textbf{(iii)} access to external resources such as \texttt{C}-code,
processes, etc. Our problem is to some extent similar to porting
\texttt{C}-programs between different compilers and operating systems.
Although today's \texttt{C}-environments have made significant
progress in standardising the structure of the library
(e.g., \texttt{C99} internationalisation support) and POSIX has greatly
simplified operating system portability, writing portable
\texttt{C}-code still relies on judicious use of the
\texttt{C}-preprocessor and a principled approach to
portability. We will take advantage of the underlying principles
and choices that affect portability in the \texttt{C}-world, both
because we believe the examples are widely known and because the
\texttt{C}-community has a long-standing experience with portability issues.
Note that the described approaches are not mutually exclusive.

\paragraph{The abstraction approach}

A popular approach to make an application portable is to define an
\emph{interface} for facilities that are needed by the application and
that are typically not portable.  Next, the interface is implemented
for the various target platforms.  Targets that are completely
different (e.g. Windows vs. X11 graphics) use completely distinct
implementations, while small differences are handled using
compile-time or run-time conditions.  Typically, the ``portable'' part
of the application still needs some conditional statements, for
example if vital features are simply not available on one of the target
platforms.

Abstractions come in two flavours: specifically designed and
implemented in the context of an application; and designed as
high-level general-purpose abstractions. We find instances of the
latter class notably in areas where portability is hard, such as
user-interface components (e.g., WxWindows, Qt, various libraries for
threading).

Logtalk~\cite{pmoura03} is an example from the Prolog world: it
provides a portable program-structuring framework (objects) and
extensive libraries that are portable over a wide range of Prolog
implementation.  On the other hand, we could claim that Logtalk is a
\emph{language} developed by a community that just happens to be using
a variety of Prolog implementations as backend.  The portability of
Logtalk itself is based on application-specific abstraction.

\paragraph{The emulation approach}

Another popular approach is to write applications for environment $X$
and completely \emph{emulate} environment $X$ on top of the target
environment $Y$. One of the most extreme examples here is
\emph{Wine}\urlref{http://www.winehq.org}, that completely emulates the Windows-API on top
of POSIX systems.  The opposite is Cygwin~\cite{cygwin}, that emulates the
POSIX API on Windows platforms.

This approach has large advantages in reducing the porting effort.
However, it comes at a price. Cygwin and Wine are very large projects
because emulating one OS API is approaching the complexity of an OS
itself. This means that applications ported using this approach become
heavyweight. Moreover, they tend to become slow due to small mismatches.
For example, both Windows and POSIX provide a function to enumerate
members of a directory and a function to get details on each member. The
initial enumeration already provides more than just the name, but the
set of attributes provided differs. This implies that a full emulation
of the directory-scanning function also needs to call the `get-details'
function to fill the missing attributes, causing a huge slow-down. The
real pain is that often, the application is not interested in these
painfully extracted attributes. Similar arguments hold for the
differences between the thread-synchronisation primitives. For example,
the initial implementation of SWI-Prolog message-queues that establish a
FIFO queue between threads was based on POSIX thread `condition
variables' and ported using the pthread-win32\urlref{http://sourceware.org/pthreads-win32/} library. The Windows
version was over 100 times slower than the POSIX version.  Rewriting
the queue logic using Windows `Event' object duplicates a large
part of the queue-handling code, but provides comparable performance.

\paragraph{The conditional approach}

Traditionally, (small) compatibility problems are `fixed' using
conditional code.  There are two approaches: compile-time and
run-time.  In the Prolog-world, we've seen mostly run-time solution
with the promise that partial evaluation can turn this into the
equivalent of the compile-time approach.

Conditions themselves often come from version information (e.g.\ if (
currentBrowser == IE \&\& browserVersion == 6.0 ) ...). At some point
in time, the variation in the Unix-world was so large that this was no
longer feasible. Large packages came with a configuration file where
the installer could indicate which features where supported by the
target Unix version. Of course, most system managers had no clue. A
major step forward was GNU \texttt{autoconf}~\cite{citeulike:130549}, a
package that provides clear guidelines for portability, plus a
collectively maintained suite of tests that can automatically execute
in the target environment (\texttt{configure}).

There is one important lesson to be learned from GNU autoconf:
\textit{do not test versions, but features}. E.g.\ if you want to know
whether \index{member/2}\predref{member}{2} is available without loading library(lists), use a test
like the one below rather than a test for a specific Prolog
implementation and version. Feature-tests like this are the basis of
autoconf.  Where autoconf requires writing an m4 specification file
that is translated into the well-known configure program and the test
results must be queries using \verb$#ifdef HAVE_$\bnfmeta{function}, the
reflexive capabilities of Prolog avoid the need for external toolchains.

Feature tests work regardless of your knowledge of the availability of
a predicate in a specific Prolog implementation and they keep working
if implementations change this aspect or new implementations arrive on
the market.

{\footnotesize
\begin{verbatim}
    catch(member(a, [a]), _, fail)
\end{verbatim}}

\noindent
%================================================================
\section{Prolog portability status}

Before we can answer the question on the best approach for Prolog, we
must investigate the situation. The relevant situation does not only
include the target Prolog systems, but also the user and developer
communities.

Our target Prolog systems have been influenced by the Edinburgh
tradition, namely through Quintus Prolog, \texttt{C}-Prolog, DEC10-Prolog
and its DEC10 Prolog library. They all support the ISO core
standard. In addition, resources such as Logtalk, and the Leuven and
Vienna constraint libraries have recently helped enhancing the compatibility
of Prolog dialects due
to a mutual interest of the resource developers (a wider audience) and
Prolog implementors (valuable resources). Logtalk has pioneered this
field, pointing Prolog implementors at non-compliance with the ISO
standard and other incompatibilities. The constraint libraries have
settled around the attributed variable and global variable API
designed for hProlog (\cite{Demoen:CW350}). These APIs are either directly
implemented or easily emulated.

\begin{table}[b!]
\def\N{\footnotemark}
\renewcommand{\thefootnote}{\textit{\alph{footnote}}}
\begin{minipage}{\textwidth}
\begin{center}
\begin{tabular}{|l|c|c|c|c|}
\hline
&				Ciao	& SICStus    & SWI	& YAP \\
\hline
ISO			      &	yes	& yes	     & yes	& yes \\
\index{module/2}\predref{module}{2}		      &	yes	& yes	     & yes\N[1]	& yes\N[1] \\
\index{module/3}\predref{module}{3}		      &	yes	& no	     & no	& no \\
\index{use_module/2}\predref{use\_module}{2}		      &	yes	& yes	     & yes	& yes \\
\index{use_module/3}\predref{use\_module}{3}		      &	no	& yes	     & no	& no \\
operators and modules	      &	local	& global     & both	& both \\
export built-in		      &	no	& no	     & yes	& yes \\
redefine built-in	      &	yes	& no	     & yes	& yes \\
\hline
Term-expansion		      &	yes	& yes	     & yes	& yes \\
Goal-expansion		      &	yes	& yes	     & yes	& yes \\
Compilation-model\N[2]	      &	file	& direct     & direct	& direct \\
Directives		      &	special	& goal       & goal	& goal \\
\hline
Attributed variables	      &	yes	& yes	     & yes	& yes \\
Coroutining (\index{dif/2}\predref{dif}{2}, \index{freeze/2}\predref{freeze}{2}) &	yes	& yes	     & yes	& yes \\
Global variables	      &	yes	& yes	     & yes	& yes \\
Tabling			      &	yes	& no	     & no	& yes \\
Threads			      &	yes	& no	     & yes\N[3]	& yes\N[3] \\
Unicode			      &	no	& yes	     & yes	& yes\N[8] \\
\hline
Set unknown flag	      &	fail	& error	     & yes\N[4]	& yes\N[4] \\
Get unknown flag\N[5]	      &	fail	& fail	     & fail	& fail \\
Provide unknown option\N[6]   &	error	& error	     & ignore	& error \\
\hline
Library license		      &	GPL	& Proprietary & GPL\N[7]& Artistic \& GPL \\
\hline
\end{tabular}
\footnotetext[1]{Allows exporting operators}
\footnotetext[2]{File: compile .pl to object and load object code}
\footnotetext[3]{Provides \index{create\_prolog\_flag/3}\predref{create\_prolog\_flag}{3}}
\footnotetext[4]{Following ISO technical report}
\footnotetext[5]{TBD: Doesn't ISO demand an error?}
\footnotetext[6]{E.g.\ write\_term(foobar, [hello(true)])}
\footnotetext[7]{With an additional statement that allows for use in
		 proprietary code, based on the GCC runtime library.}
\footnotetext[8]{Only at the scanner level.}
\end{center}
\end{minipage}

    \caption{Core features provided by the target Prolog environment}
    \label{tab:features}
\end{table}

\paragraph{The language}

All systems can run programs satisfying the ISO standard as long as they
do not depend on corner cases. There are cases where ISO demands an
exception and implementations take the liberty to provide meaningful
semantics. E.g., SWI-Prolog supports the mode \term{arg}{-,+,?}; many
systems support `options' to predicates such as \index{open/4}\predref{open}{4} and \index{write\_term/4}\predref{write\_term}{4}
that are not described by the ISO standard (e.g. `encoding' in \index{open/4}\predref{open}{4} to
indicate the character-set encoding of the file). Additional options are
explicitely allowed by the standard, but there is no good mechanism to
know which options are allowed by a specific implementation and it is
not easy to find an elegant way to deal with different option-list
requirements in different implementations. Similarly, most systems
provide prolog-flags (\index{current_prolog_flag/2}\predref{current\_prolog\_flag}{2}) in addition to the standard
flags. Finally, systems differ in the relation between operators and
modules. \Tabref{features} provides an overview of relevant features in
the four Prolog dialects considered.\footnote{See also
\url{http://en.wikipedia.org/wiki/Comparison\_of\_Prolog\_implementations}}

\paragraph{The library}

The situation around the Prolog libraries is unfortunate. Although
much of the code is derived from the public domain `DEC10' library, a
long period of independent development makes this barely
recognisable. The aforementioned cooperation around Logtalk and the
CLP libraries as well as discussions in Leuven\urlref{http://www.cs.kuleuven.ac.be/~dtai/projects/ALP/newsletter/feb09/content/Articles/doc5/content.html} have enhanced the
situation somewhat.  Reaching compatibility by
re-mixing a new library from all available libraries involves reaching
agreement on structure (e.g., files and modules), predicate names and
semantics and resolving license issues.

Currently, the way predicates are spread over the libraries and system
built-ins differs enormously. Also different is the status of built-in
predicates (can you redefine them, can you export them from a library,
etc.) differs.  Fortunately, there are only few cases where we find
predicates with the same name but different semantics
(e.g. \index{delete/3}\predref{delete}{3}\urlref{http://www.cs.otago.ac.nz/staffpriv/ok/pllib.htm})

\paragraph{Foreign code}

As Bagnara (\cite{bagnara:02a}) points out, the design of the foreign
language interface is largely settled. All target systems use
`term-handles'; opaque handles to Prolog terms that must be allocated
and thus ensure that the Prolog engine knows which terms are referenced
by foreign code. Details, such as the naming, coverage of the API
functions to interact with terms as well as the way foreign code is
made visible as Prolog predicates vary.  We identify two problem areas.

\begin{itemize}
    \item All Prolog systems allow binding external I/O channels to
    Prolog streams.  The design of these interfaces however differs
    so widely that emulation is non-trivial and likely to cause
    severe performance degradation.  See \secref{yapforeign}.

    \item The SWI and YAP APIs allow for creating non-deterministic
    predicates in C.  SICStus and Ciao require the non-determinism
    to be moved to Prolog. It is hard to make a SWI/YAP
    non-deterministic implementation run of SICStus/Ciao without
    significant rewriting.
\end{itemize}

\paragraph{Community issues}

Both the user and developer communities around Prolog are small and
flexible units. This is important to note, because as we have seen in
\secref{approach}, full emulation often becomes hard due to small
semantic differences.  The Prolog community is sufficiently flexible
to provide workarounds, as long as the impact on a system that needs
changing is minimal.\urlref{http://groups.google.com/group/comp.lang.prolog/msg/25af4e01de8a363c}

%================================================================
\section{What approach should we use for Prolog?}

The most desirable ultimate situation is of course a well standardised
core with a comprehensive common library. However, getting agreement on
such a library and proper implementations for all platforms is not
trivial. Even if this library finally exists, it will surely attract
developers but still there are a lot of legacy applications where a
complete rewrite to the new common library is not going to happen soon.
A common library will also be based on an intersection of the target
system capabilities, leaving many legacy application partly unsupported.
How do we support applications \emph{now}?

As already pointed out in the abstract, our aim is to run the same code
on multiple Prolog systems and not to \emph{migrate} code. Provided that
all required core features (see \tabref{features}) are supported in the
target system, migrating code is generally fairly easy.

As far as we are aware, there are none or very few cases where emulation
leads to poor performance due to mismatches in the APIs as explained in
\secref{approach}. So, as a good shared abstraction is hard to achieve
and application-abstractions are too limited in scope for our purposes,
\emph{emulation} is the most promising route to follow. Note that, given
a good framework, an emulation layer can be established incrementally
and on `as needed' basis.

%================================================================
\section{The need for macro-expansion.}

Certainly, we need some form of macro-processing. Dealing with
incompatibilities only using runtime tests and optionally partial
evaluation is insufficient. First of all, runtime tests can only deal
with predicates and not with declarations (directives). Second, portable and adequate
partial evaluation is not provided. Without partial
evaluation, runtime testing is not acceptable for time-critical code and
static analysis tools will complain about the code intended for other
dialects. Term- and goal-expansion are provided by all target systems,
but the details vary, making it rather awkward to use in application
code. For example, Ciao requires special attention to make the rules
available to the compiler. SWI-Prolog expansion follows its
module-inheritance rules, first expanding in the module, then in the
\const{user} module and finally in the \const{system} module. SICStus provides
additional arguments to deal with source-locations, etc.

Following the emulation-approach, compatibility libraries can use all
machinery available to the hosting Prolog environment to emulate the
target. What is still needed is something to achieve portable
conditional compilation in the application. Portable conditional
compilation remains necessary to provide a partial port if the target
lacks certain features (e.g., if the target is lacking unicode support
it might still be possible to achieve a useable application). Sometimes,
features of one system allow for realising a better
(e.g., faster, more compact) implementation for a certain subsystem.
For example, SWI-Prolog's \index{nb\_setarg/3}\predref{nb\_setarg}{3} allows for a clean reentrant
and thread-safe implementation of counting proofs that is faster and
requires less space than portable solutions.  We can code this as
below.

{\footnotesize
\begin{verbatim}
:- meta_predicate proof_count(0, -).
:- if(current_predicate(nb_setarg/3)).
proof_count(Goal, Count) :-
        State = count(0),
        (   call(Goal),
            arg(1, State, C0),
            C1 is C0 + 1,
            nb_setarg(1, State, C1),
            fail
        ;   arg(1, State, Count)
        ).
:- else.
proof_count(Goal, Count) :-
        findall(x, Goal, Xs),
        length(Xs, Count).
:- endif.
\end{verbatim}}

\noindent
%================================================================
\section{The SWI/YAP portability framework}

The SWI/YAP approach is based on emulation.  Its key features are:

\begin{itemize}
    \item Support \verb$:- if(Goal).$  \ldots
		  [\verb$:- else.$ \ldots ]
		  \verb$:- endif.$ conditional compilation.  This
    is now built-in SWI and YAP, but can easily be provided on top
    of term-expansion for other systems.

    \item Provide \verb$:- expects_dialect(Dialect).$ to state that
    a module is designed for the given dialect.  The effect of this
    directive is threefold.

    \begin{enumerate}
	\item Load and import library(dialect/Dialect), which provides
	emulation for built-ins of the dialect and term/goal expansion
	rules to resolve compatibility issues.

        \item Make the current dialect available through
	\term{prolog\_load\_context}{dialect, Dialect} for term and
	goal-expansion.

	\item Push a new library directory before the current library
	path.  The new directory can provide additional and replacement
	libraries that provide the interface of the target and use the
	implementation techniques of the host.
    \end{enumerate}

    \item Synchronise some vital features, such as identifying the
    running dialect using the Prolog flag \const{dialect}.

    \item Provide a C-header to emulate the target foreign interface.
    Given the similar design, the header consists mostly of typedefs,
    macros to deal with simple renaming and a few (inline) functions
    for more complicated cases.
\end{itemize}

%================================================================
\section{Making SWI-Prolog foreign resources available in YAP}
\label{sec:yapforeign}

YAP emulates a large subset of the SWI-Prolog interface library. This
was largely done in order to facilitate porting of SWI-Prolog
applications to the YAP system. Currently, YAP emulates the main
functionality in the SWI interface, and YAP has been able to run complex
SWI applications that heavily use \texttt{C}-code, such as the
\texttt{jpl} java-prolog interface, and the \texttt{sgml} library.

YAP implements around $140$ functions in the SWI-interface, altogether
they require over 2000 lines of code.\footnote{The interface contains
significant duplicate functionality because old functions have been
replaced by more powerful ones. For example, strings were originally
exchanged as 0-terminated C-strings, then using an additional length
parameter to accommodate 0-bytes in atoms before reaching the current API
that accepts a flag parameter to represent types (PL\_ATOM, PL\_STRING)
and encoding (REP\_ISO\_LATIN\_1, REP\_UTF8), etc. Ideally, the deprecated
functions should be marked as such and be provided as macros mapping to
the new API.} The interface originally was implemented as a layer over
the native YAP C-interface, but more recently, we have decided to
integrate the SWI-Prolog interface as an alternative to the native
interface. This avoids the cost of going through two layers of
interfacing. Next, we discuss the main challenges we had to address in
our implementation.

The first challenge is sheer size: SWI-Prolog exports over $200$
functions. Implementing the whole functionality in a single go would
have been a major endeavour. Instead, we chose to implement functions
as they are needed by the applications we need to port. The
one-step-at-a-time approach was also used to implement complex
interface functions. This is risky: we have to be careful to inform
users that an interface function is only \emph{partially} implemented.

The second challenge were the differences in internal objects that
were exported through the interface. As an example, SWI-Prolog
internally supports an integer Prolog object that is always 64 bits
long. YAP supports an integer that has word size. This creates a
problem in 32-bit machines, as 64-bit integers have to be processed as
big numbers. As a second example, SWI-Prolog supports a string object:
YAP does not support such objects, instead strings are processed to
lists of character codes. In practice, strings are not very popular in
the applications we experimented with.

We did observe major differences in functionality between the two
systems. Notice that from the YAP point of view, this is a problem when
SWI-Prolog has functionality that does not exist in YAP. One example is
the debugging infrastructure, that is much richer in SWI-Prolog. A
second typical example are \textit{blobs}. A blob is a symbol (like an
atom) that is used to store external data, such as image-pixels or a
handle to C-managed data. YAP uses blobs and has some support for
different types of blobs. SWI-Prolog goes much further, and has a
sizeable infrastructure for blobs that accommodates user defined blobs
with extensions over input, output, garbage-collection, etc. In cases
such as this, supporting the SWI interface will dictate how blobs will
be supported in YAP. The advantage is that YAP will benefit from the
decisions made by SWI. The drawback is that the YAP design is bound by
these decisions.

The third challenge is in Input/Output. SWI-Prolog basically exports
its Input/Output data structures, which are very different from
YAP's. A first try at using the standard emulation layer approach was
very painful: first because the interface is complex; and second
because it involves reimplementing a large number of
data-structures that had to be working before anything could be
experimented with. On the other hand, we could observe that
SWI-Prolog's I/O was largely self-contained and almost exclusively
written in \texttt{C}.  This suggested an alternative approach, where
it was decided to simply port the whole I/O subsystem as a \texttt{C}
library. The process worked surprisingly well: the I/O routines are
much independent of the rest of the system, and we only required
reimplementing some internal interface functions. The interface layer
require 800 lines of code, but much of this code is in fact reused
from files in SWI-Prolog. We did observe two difficulties:

\begin{itemize}
\item some I/O functions build lists of characters using low-level
  abstract machine functionality; we just abstracted these operations
  without loss of efficiency.
\item the code relies on the value of some atoms being known at
  compile-time. This is currently not supported in YAP, so the
  initialisation had to be implemented at run-time in YAP.
\end{itemize}

The one major problem is that now YAP has two independent I/O
routines: the SWI and the original ones. Ultimately, either YAP should
support only the SWI ones, or we will need to allow both to coexist
gracefully. In either case, it is a hard decision.

The last challenge is simply keeping track of the changes in SWI
functionality.  SWI-Prolog is a living object: new
functions are being added in, and from time to time, preexisting functions
do change. This is a good thing, and just a small problem with the external interface, but it
is a major problem with the I/O library. As YAP-6 stabilises, we expect
to be able to merge the YAP changes to the main SWI distribution, and
use \texttt{git} to track down changes in the SWI distribution,
with no negative impact on SWI.

%================================================================
\section{Portable constraint libraries}

We have been able to share three major constraint libraries between the
two systems using this framework:
\texttt{clpfd}~\cite{DBLP:conf/iclp/Triska08},
\texttt{clpr}~\cite{DBLP:conf/plilp/Holzbaur92}, and
\texttt{chr}~\cite{DBLP:series/lncs/KoninckSD08}. YAP originally
implemented a SICStus mechanism for domain variables, so the first step
was to also support the hProlog/SWI-Prolog mechanism~\cite{Demoen:CW350}. The main
differences are:

\begin{itemize}
\item SICStus requires some preprocessing, as the possible attributes
  must be specified at compile-time;
\item SICStus allows an extra-step, after query execution and before
  creating the constraint goals, which is not available in the hProlog
  design.
\item SICStus often provides access to all the attributed
  variables. This is useful for simplifying the global constraint-store.
  It is also not clear whether it should (or not) include attributed
  variables in global variables.
\end{itemize}
Arguably, the hProlog interface can be seen as more ``lower-level'' than
the SICStus approach. From YAP-6.0.4, YAP implements the SICStus
interface as mostly an extension of the SWI interface (with some
extra built-ins). Following SWI-Prolog, YAP now simply searches the
global stack for attributed variables for realising \index{call_residue_vars/2}\predref{call\_residue\_vars}{2},
which is used by the toplevel to report residual constraints.

Given a common infrastructure, the goal was to reduce to the least the
amount of effort in porting the constraint libraries between the two
different systems. In the case of \texttt{chr} this was simplified
because \texttt{chr} already supported two systems: SICStus and
SWI. Difficulties had to do with the term expansion mechanism, which
is different in the two systems, with SWI-Prolog having a more liberal
syntax, and with supporting SWI's \textit{message-writing}
mechanism.\footnote{Based on Quintus Prolog. See \index{print_message/2}\predref{print\_message}{2}.} Last,
chr was originally implemented in \texttt{hProlog} and expects an
\texttt{hProlog} compatibility library to provide list functionality.
This forces YAP to be both compatible with SWI and hProlog.

Markus Triska's \texttt{clpfd} is a SWI-native application. It was
interesting that although the two applications were written
independently, the challenges were very much similar: the term expansion
mechanism, using the message-writing system, and attribute predicates.

%================================================================
\section{A case-study: the Alpino dependency-tree parser suite}

The Alpino dependency-tree parser suite \cite{vannoord-taln} is a large
and complicated program developed in SICStus Prolog over a long period
of time. \Tabref{alpino} gives some metrics of the application. The
initiative to port Alpino came from the SWI-Prolog side based on a
desire to use Alpino components as a library in a larger SWI-Prolog
based application. On first contact, the Alpino team was interested, but
had two major worries: ``does SWI-Prolog support our current application
without major rewrites'', and ``can we achieve one source that compiles
and runs on both''. The first was accompanied with a list of
requirements. Most of these could be answered positively without
hesitation. SWI-Prolog however lacks \index{call_residue/2}\predref{call\_residue}{2} and a Tcl/Tk
interface. SWI-Prolog has a partial implementation of
\index{call_residue_vars/3}\predref{call\_residue\_vars}{3}.\footnote{The implementation may report variables
that are inaccessible due to backtracking if the application uses
non-backtrackable assignment as defined by \index{nb_setarg/3}\predref{nb\_setarg}{3} and \index{nv_setval/2}\predref{nv\_setval}{2}.}
Later \index{copy_term/3}\predref{copy\_term}{3} proved the correct and portable solution for the
application's purposes. Tcl/Tk was no hard requirement and we hoped that
the Ciao implementation might be able to solve this issue. A short
summary of the SWI/YAP portability framework convinced the Alpino team
that future maintenance based on a common source could de dealt with.

\begin{table}
\begin{center}
\begin{tabular}{ll}
\hline
Prolog source-files & 304 \\
Prolog source-lines & 473,593 \\
Prolog predicates & $\pm$ 5,500 \\
Prolog clauses & $\pm$ 290,000 \\
C source-files & 14 \\
C++ source-files & 27 \\
C/C++-defined predicates & 46 \\
\hline
\end{tabular}
\end{center}
    \caption{Metrics on the Alpino Parser}
    \label{tab:alpino}
\end{table}

Below we summarise the non-trivial issues encountered and their resolution.

\begin{itemize}
    \item The SICStus block directive declares predicates to suspend
    until an instantiation pattern is reached.  SWI-Prolog
    has no such concept.  Term-expansion was used to rename the clauses
    and generate a wrapper that implements the coroutining using \index{when/2}\predref{when}{2}.%
    \footnote{Eventually, it was decided that using \index{when/2}\predref{when}{2} directly was
    more elegant and natively supported by both target Prolog systems.}

    \item Operator declarations are mapped to declarations in the user
    module, SWI-Prolog's deprecated support for system-wide operators.
    The code below illustrates dialect handling here:

    \footnotesize{
    \begin{verbatim}
system:goal_expansion(op(Pri,Ass,Name),
                      op(Pri,Ass,user:Name)) :-
        \+ qualified(Name),
        prolog_load_context(dialect, sicstus).

qualified(Var) :- var(Var), !, fail.
qualified(_:_).
    \end{verbatim}}

\noindent
    \item Alpino depends on predicates from library(lists) that
    we do not consider for including into SWI-Prolog.  Therefore,
    we add library(dialect/sicstus/lists) with the following content

\footnotesize{
\begin{verbatim}
:- module(sicstus_lists,
          [ substitute/4,       % +Elem, +List, +NewElem, -List
            nth/3
          ]).
:- reexport('../../lists').

<implementation>
\end{verbatim}}

\noindent
    Note that in addition, we must map explicitly qualified calls
    (e.g., lists:nth(N,L,E)) to sicstus\_lists:nth(N,L,E) if the current
    dialect is sicstus.  The mapping rule is in \file{sicstus.pl}, while
    clauses for the mapping are provided by the renamed modules.

    \item database references (\index{assert/2}\predref{assert}{2}, \index{clause/3}\predref{clause}{3}, \index{recorda/3}\predref{recorda}{3}, \index{erase/1}\predref{erase}{1})
    are safe in SICStus and goals fail if the reference does not exist.
    SWI-Prolog references used to be unsafe: references were heuristically
    tested for validity and an existence\_error was raised if the reference
    was known to be invalid.  In case the heuristics incorrectly claims
    that a reference is valid, the system could crash.  Programming
    around this in Alpino was considered more effort than providing a
    compatible API in SWI-Prolog, so we decided for the latter.\footnote{The
    necessary infrastructure was developed several years ago.}

    \item We added support for the mode \term{recorded}{-,+,-} to the
    SWI-Prolog runtime.  We also resolved that \bnfmeta{m}:clause(H,B) does not
    qualify \arg{H} if the predicate is in module \bnfmeta{m}.

    \item SICStus (and Ciao) provide Prolog streams that can both the
    read and written to.  SWI-Prolog's streams are either read or
    write.  This makes it hard to provide a compatible emulation of
    the sockets library.  We decided to support stream-pairs in the
    SWI-Prolog runtime system.  All I/O predicates are aware of these
    pairs and will pick the appropriate member (\index{close/1}\predref{close}{1} addresses both
    streams).  After this addition, emulating the required features of
    the socket library was simple.

    \item SICStus assert and friends can deal with attributed variables,
    as illustrated below.

\footnotesize{
\begin{verbatim}
?- dif(X, 3), assert(not_3(X)).
\end{verbatim}}

\noindent
    SWI-Prolog has no such support and adding this is a non-trivial
    exercise.  As a work-around, we use goal-expansion to map calls
    to the assert-predicates onto clp\_assert.  This predicate uses
    \term{copy\_term}{+Attributed, -Plain, -Constraints} to extract
    the constraints from the term and inserts all constraints at
    the start of the body, creating the clause below.

\footnotesize{
\begin{verbatim}
not_3(X) :- dif(X, 3).
\end{verbatim}}

\noindent
    We consider the approach so specific that we decided to make the
    emulation part of the Alpino source-tree rather than the
    SWI-Prolog system.

    \item We provide an implementation for the libraries \file{arrays.pl},
    \file{system.pl} and \file{timeout.pl} using SWI-Prolog primitives.

    \item At some places, we decided that both SICStus and SWI-Prolog
    provided already compatible alternatives for legacy SICStus code
    and adjusted the Alpino sources accordingly.

    \item We emulate the declaration of foreign predicates using
    the SICStus primitives \index{foreign_resource/2}\predref{foreign\_resource}{2}, \index{foreign/3}\predref{foreign}{3} and load
    \index{foreign_resource/1}\predref{foreign\_resource}{1}.  The wrapper-generation is an extension of
    the older generator for Quintus (qpforeign.pl).  In addition we
    wrote a script emulating the features of \program{splfr} that we
    need.  This SICStus program extracts the foreign declaration from
    a Prolog file, generates a wrapper and calls the C-compiler to
    create a loadable foreign module.  The SWI-Prolog replacement
    \program{swipl-lfr.pl} takes the same steps, using the C-compiler
    and linker front-end \program{swipl-ld} for the platform-specific
    linking.

    In addition, we added sicstus.h to the SWI-Prolog include directory
    that provides the necessary mapping from SP\_* API functions to
    PL\_* API functions.  The total amount of code involved is 664 lines
    of Prolog code and 244 lines of C-header (which satisfies our requirements,
    but is otherwise incomplete).  No changes were required to the Alpino
    C-files, neither to the Prolog code.  For the Alpino zlib-interface,
    creating a compressed serialisation of a Prolog term based on SICStus
    fastrw.pl library and zlib, we decided on an alternative route for SWI
    that was easier to realise than providing fastrw for SWI-Prolog.  The
    Alpino code selects the implementation using the \index{if/1}\predref{if}{1} conditional
    compilation.

    \item Alpino uses the SICStus tcl/tk interface.   License issues
    make it impossible to use the SICStus library here, while reimplementing
    from scratch is non-trivial. Initially, we ported library(tcltk)
    from Ciao Prolog using the same emulation-approach. Because Ciao
    uses a much finer grained module infrastructure, emulating enough
    of Ciao to run the tcltk library requires 17 files containing 971
    lines of Prolog.  In addition, SWI-Prolog's \index{write_term/3}\predref{write\_term}{3} had to
    be modified to (by default) omit an extra space after a comma that
    separates two arguments (e.g., \verb$term(a,b)$ instead
    of \verb$term(a, b)$).\footnote{This issue also affected Alpino,
    which contains C-code that relied on the exact term-layout. The
    13211-1 standard describes spaces in the output of write\_term to
    separate tokens where needed. Other spaces are not
    \emph{explicitly} forbidden.}

    Unfortunately, Ciao's tcltk library could not sufficiently emulate
    the SICStus library for running Alpino.  Eventually, the Ciao code
    was used to realise a new and portable tcl/tk interface that could
    support Alpino.  This interface is part of the Alpino source-tree.
\end{itemize}

The above changes required about 20 person-days joined effort from the
SWI-Prolog team and the Alpino team and resulted in a fully operational
application running on the two target platforms. As mentioned above,
SWI-Prolog was enhanced in several places. Also the Alpino code has been
improved. It now relies less on SICStus legacy code; the application now
supports UTF-8 on both Prolog platforms; the modularity was enhanced and
the performance has been improved, also on SICStus.

The initial Alpino source contained 19 places of conditional compilation
based of the \index{if/1}\predref{if}{1}-directive. Since then, more conditional code was added
to enhance performance on SWI-Prolog and use additional features of
SWI-Prolog, such as (partial) support for multi-threading and its
interface to GNU readline. The current code contains 59 places of
conditional compilation. This small amount of conditional code has no
significant impact of the maintainability of the Alpino code-base.

%================================================================
\section{Conclusions}

Portability of Prolog source-code is important. Portability prevents
vendor lock-in, provides backup if an implementation is discontinued or
is no longer suitable for sustaining an application because it lacks
features that are important for future development. Portability is also
needed if we want to combine packages developed on different Prolog
implementations. For a long time, the Prolog community consisted of
separated sub-communities associated to an implementation. The ISO
standard has resolved many low-level compatibility issues. Logtalk and
the Leuven/Vienna constraint libraries have created bridges, causing
participating Prolog systems to resolve various incompatibilities.
Currently, portability among four systems with common inspiration (YAP,
SICStus, Ciao and SWI-Prolog) is comparable to other multi-vendor
programming environments such as C on Unix in the 90s.

We have presented a framework that provides conditional compilation,
where the reflexive capabilities of Prolog replace the analysis provided
by GNU autoconf. We also presented a framework that allows sources for
multiple dialects to co-exist on the same Prolog host. This framework
can be extended on `as-needed' basis.

We identified a number of issues that hinder the development of
portable Prolog resources. Some of these involve major decisions and
require major effort. Examples are non-portable types such as
string-objects, advanced numeric types (unbounded, rationals,
complex), and non-portable features (e.g., Unicode support, threads,
tabling). There are a number of issues that are less involved and can
greatly facilitate portability if agreement is reached and
implemented. Examples are `environment predicates', such as
\index{absolute_file_name/3}\predref{absolute\_file\_name}{3}, \index{prolog_load_context/2}\predref{prolog\_load\_context}{2}, a mechanism to deliver
(translated) messages to the user, further standardisation of Prolog
flags, including a mechanism to define new flags and a clear vision on
handling extensions to the option-list processed by predicates such as
\index{write_term/3}\predref{write\_term}{3}.

We strongly advice anyone interested in porting a Prolog resource to get
into contact with the vendors of the targeted Prolog systems. Many
incompatibilities are much easier resolved by the vendor(s) and as a
result both systems improve and get more compatible.

% \bibliographystyle{plain}
% \bibliography{pl,pldoc}

\end{document}